# Status of the Beryllium Tile Bonding Qualification activities for the Manufacturing of the ITER First Wall

Raphaël Mitteau[a], R. Eaton[a], G. Perez[a], F. Zacchia[b], S. Banetta[b], B. Bellin[b], A. Gervash[c], D. Glazunov[c], J. Chen[d]

[a]ITER Organization, Route de Vinon-sur-Verdon, CS 90 046, 13067 St. Paul Lez Durance Cedex.
[b]Fusion for Energy, Josep Pla 2, Torres Diagonal Litoral B3, 08019, Barcelona, Spain
[c]Efremov Research Institute, 189631, St. Petersburg, Russia
[d]Southwestern Institute of Physics, Huangjing Road, Chengdu 610225, China

The preparation of the manufacturing of the ITER first wall involves a qualification stage. The qualification aims at demonstrating that manufacturers can deliver the needed reliability and quality for the beryllium to copper bond, before the manufacturing can commence. The qualification is done on semi-prototype, containing relevant features relative to the beryllium armour (about 1/6 of the panel size). The qualification is done by the participating parties, firstly by a manufacturing semi-prototype and then by testing it under heat flux. One semi-prototype is manufactured and is being tested, and further from other manufacturers are still to come. The qualification program is accompanied by bond defect investigations, which aim at defining defect acceptance criteria. Qualification and defect acceptance program are supported by thermal and stress analyses, with good agreement regarding the thermal results, and some insights about the governing factors to bond damage.

ITER, First Wall, Qualification, prototypes, high heat flux tests.

## 1. Introduction

The first wall (FW) is the main protection of the ITER vessel against the plasma. The first wall is made of 440 panels with typical dimensions of 1.5 m toroidally and 1 m poloidally, cladded with 8-10 mm thick beryllium (Be) armour tiles of 16 to 42 mm in size [1]. The first wall design is based on toroidal plasma facing units (the "fingers"), which are cantilevered to a structural beam oriented in the poloidal direction. The beam is the structural backbone to the first wall panel, serving both as a mechanical supporting structure and containing all the needed interfaces to the shielding blocks (water connection and distribution to the fingers, electrical earthing, remote-handling compatible attachment, See fig.2 in Ref 1). Two series of fingers on each side of the beam form two beryllium armoured wings, separated by a recessed poloidal slot giving access to the beam. The fingers have a typical length of 0.75 m. They are made of a steel / copper alloy / beryllium sandwich. The steel base of the finger has the role of supporting structure, the copper is the heat sink, and the beryllium is the armour. The water is fed through the steel, cools the heat sink, and is returned to the finger extremity through the steel base. The fingers are designed to accommodate heat fluxes from 2 (Normal Heat Flux - NHF) to 4.7 MW/m² (Enhanced Heat Flux - EHF) depending on the panel location in the first wall [1]. They are designed to 15000 heat loading cycles. The NHF fingers rely mainly on hipped bonds, including the cooling circuit made of steel pipes and also the beryllium tiles to the heat sink. The EHF fingers are made of exploded CuCrZr to steel, and the cooling circuit is completed by welding the fabricated steel base, and the beryllium tiles are assembled by brazing. The FW has passed successfully its final design review in 2013 [2]. It is now in a pre-manufacturing phase with the preparation and signature of procurement arrangements. The FW is procured by EU (50% of the panels), RF (40%) and CN (10%) Domestic Agencies (DA).

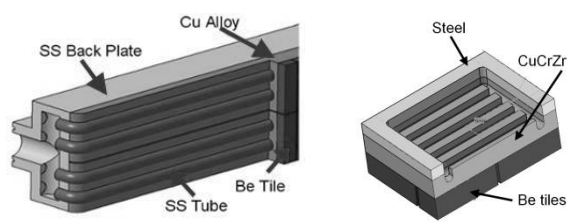

1.a NHF cooling structure of the PFC unit  　　1.b EHF cooling structure of the PFC unit

Fig. 1. Typical NHF and EHF fingers

This paper presents the qualification process that is taking place between the R&D and manufacturing phase (section 2). In addition, section 3 presents the parallel activities relative to the modelling of the bond behaviour with respect to the effects of defects and attempts at understanding the defect propagation criteria.

## 2. Qualification

The manufacturing routes toward the panel fabrication involve many challenging fabrication steps, including the bonding of the beryllium armour tiles to the copper alloy heat sink, the manufacturing of an intricate cooling circuit, integration of mechanical and hydraulic connections, isolating coatings, plus others not mentioned here. Among these delicate processes, successfully bonding the beryllium armour tile to the copper alloy heat sink is especially challenging, for two reasons : 1/ beryllium and copper have different coefficient of thermal expansion (CTE, $\Delta_{CTE} = 4.10^{-6}$) , hence large shear stresses develop in the bond when temperature changes. The stress grows very large, due to the singularity at the free edge of the bond. The bonding is solicited with temperatures changes of several hundreds of degrees firstly during the manufacturing step, then during the operation life. The tile bonding joint must withstand stresses close to or beyond the yield strength. 2/ the phase diagram contains brittle intermetallic and intermediate metastable phases [3].

There is currently no significant industrial application with beryllium to copper bonding for commercial application. The industrial experience is mainly the one built from bonding R&D which has been on-going using small scale mock-ups since more than 20 years in various fusion organisations and laboratories around the world [4-7]. Investigated topics include:

- Manufacturing processes (bonding technique, heat / pressure cycles)
- Geometrical effects (tile size, flat or curved bond, local geometry at the bond extremity)
- Material effects (base material grades)
- Intermediate layer or coating, pre assembly cleaning processes
- Effect of fast surface transient heat load and irradiation on bond properties.

This extensive R&D phase allowed to build a large knowledge basis for the behaviour of the Be/Cu bond, albeit for different base materials, tile sizes, or bonding processes. With the approach of the manufacturing phase, the R&D activity is diminishing (although some aspects are still being investigated as described later in this paper) and the focus is set on the qualification stage (Fig. 2) in preparation for the manufacture. Two bonding techniques are finally selected for joining the tiles of the ITER first wall : brazing and hot isostatic pressing [8, 9]. For these two processes, many parameters affect the bond quality. Some of them are still quite sensitive to precise manufacturing conditions, which makes them critical manufacturing steps. Part of the monitoring rely on a qualification process.

Before full scale qualification, an intermediate qualification step is defined. This qualification (sometimes called "pre-qualification") is explicitly planned by the ITER agreement for critical components being manufactured by multiple DAs. For the first wall program, this step aims at validating the tile bonding process selected by the DA and contributing to stabilise the data obtained during the R&D phase. It focuses the component development effort toward a limited number of series relevant manufacturing processes. The complete qualification occurs at the stage of the Full Scale Prototypes (FSP), and has also the objective of validating the manufacturing lines by the industrial suppliers. This final qualification is not developed here.

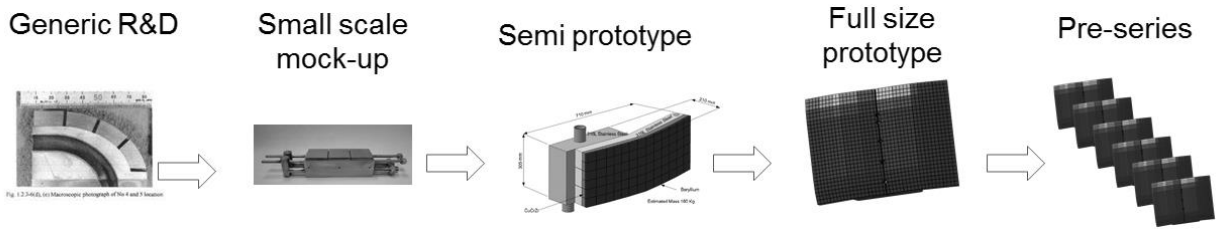

Figure 2 : First wall development route

The pre-qualification program is based on small scale mock-ups, and a defined Semi-Prototype design (SP). Both small scale mock-ups and the SP are to be tested under relevant heat load for a significant number of heat cycles.

1/ The qualification component (the SP) needs to demonstrate adequate bonding of a representative set of plasma facing unit / tiles of the final design (tile size, facets, significant number of tiles, several fingers, cooling circuit). The design of the SP is about 1/6 of the FW panel size, representing 6 plasma facing units, 3 facets, 80 – 1000 tiles depending on the tile size. The cooling circuit and finger cross-section are representative of the FW panel design. However, it is admitted that the supporting structure and component water connection might differ from the panel design. These are not critical features of the FW panel design, and are to be qualified at the FSP stage.

Table 1. Criteria for SP qualification

NHF – design heat load 2 MW/m²

| Testing step | $N^{ber}$ of test cycles | Surface to be tested |
|---|---|---|
| 2 MW/m² | 7500 | 50% |
| 2.5 MW/m² | 1500 | 5% |

EHF– design heat load 4.7 MW/m²

| Testing step | N$^{ber}$ of test cycles | Surface to be tested |
|---|---|---|
| 4.7 MW/m² | 7500 | 50% |
| 5.9 MW/m² | 1500 | 5% |

2/ The testing includes both factory tests (helium leak testing, water flow test, pressure test, UT) and high heat flux testing. The heat flux testing has to be done on a qualified heat flux test bed, which has already demonstrated the capability to testing similar components under similar testing protocols. The testing program encompasses the criteria of the 'design by experiment' rules (table 1). The success criteria are that no failure shall occur during the testing, as well as no detachment of any beryllium tile and no loss of beryllium material. There are also quantitative criteria based on the surface temperature of the beryllium tiles, namely no variation of the maximum surface temperature exceeding 20% (measured in °C) during the test, and no "hot spots" (location where temperature is higher than 30% (measured in °C) when compared to surrounding area). The success criteria are a compromise between setting a high standard for bond success, while ensuring that heat flux test can be achieved within acceptable time and resources. This is the reason for the step 2 testing at 125% the design heat load, which aims at doing accelerated fatigue. The heat load of 125% the design heat load is selected so that there is still an acceptable margin to critical heat flux. The success criteria are also backed by experience feedback from previous successful program involving water-cooled bonded plasma facing components [10].

Qualification activities are progressing in the three procuring agencies (Fig. 3). EU has completed one semi-prototype (shown in Fig 3.a), which has successfully passed factory acceptance tests and is now being tested under heat flux; one additional SP is being manufactured. RF and CN are also progressing in the manufacture of their semi-prototypes (Fig 3.b).

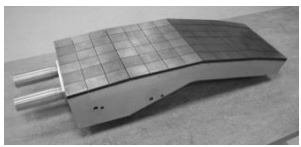    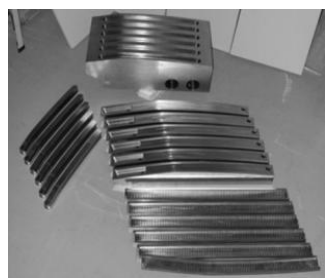

3.a : Completed NHF semi-prototype

3.b : Sub parts being prepared for a EHF semi-prototype in RF

Fig. 3. Semi-prototypes.

In parallel to the pre-qualification program, process - related R&D activities and investigations are done, aiming at a refined characterisation of the tile bonding behaviour when bonding defects are present. This is done through calibrated defects. The design size of calibrated defects has been confirmed on shear coupons, and when possible using ultrasonic testing (UT) – see figure 4. The defects are then high heat flux tested, and the defect growth is monitored using infrared camera and UT. The conventional picture is that small (< 1 mm) defects are obviously acceptable, while large (more than half the tile size) defects are certainly not acceptable. So the testing program aims at determining the critical defect size. This critical size is planned to be used as acceptance criteria during the manufacturing phase. The format of this paper does only allow to present the general outcome of these defects program. The reader is oriented to dedicated publications to have the details of each program.

For two of the manufacturing programs, large defects are not observed to evolve at the design heat flux. A large heat load (> x2 the design heat load for the NHF program) is needed to see defects evolve notably, and then the bond failure occurrence appear to be independent from the presence of a defect. At this stage, the conclusion is that defects are observed to be more resilient to propagation under heat flux cycling, compared to the previous large manufacturing experience obtained with carbon composite (CC) tiles.

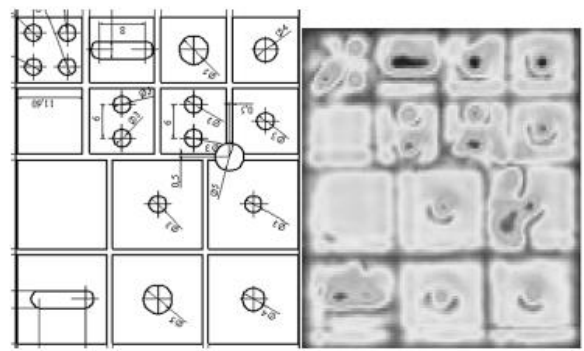

Fig. 4. EHF design bond defects and UT inspection of manufactured bond defects.

The provisional conclusion of these test programs is that the acceptable defect size shall be set by considerations other than defect growth, such as maximum acceptable temperature during operation or signal/noise ratio of non-destructive testing. The following section aims at proposing possible explanations to this favourable behaviour.

## 3. Thermal and stress analyse

Thermal and stress analyses of bond by finite element modelling (FEM) are done in support to the qualification and R&D activities [12]. A tile with a bond defect is hotter under steady state heat flux, compared to defect free tile. The excess temperature is of 40C for a 4 mm edge defect at the design heat load of 4.7 MW/m² (EHF assembly with 16 mm tile). There is an excellent agreement (usually better than 10%, one case at 25%) between measured surface temperature during the high heat flux test and the calculated temperature. This excellent agreement is a further confirmation that the planned defects are present as expected in the test mock-ups.

The stress analysis is much more delicate than temperature, because of the stress singularity at the free edge, and the absence of damage criteria for a bonded assembly [13]. While a true comparison with experiments (based on measurable quantities) is difficult to be done, the FEM stress analysis is still a helpful tool to support the interpretation of the experimental test results. Such analogy can be done by comparing trends between experiments and numerical models.

The first outcome of the stress analysis is that the stress singularity is considerably less pronounced with the copper to beryllium pair, compared to the copper to carbon pair. The stress intensity factor is typically in the range of K = 7 MPa for Cu/Be compared to 40 MPa for Cu/C, a reduction by a factor of about 8. The closer coefficients of thermal expansion (CTE) between Cu/Be (as compared to Cu/C) provide a good initial basis to explain the lower stress intensity coefficient. The smaller singularity is a good explanation to bond defects being less prone to evolve, compared to the Cu/C pair.

Moreover, the bond stress distribution at the free edge puts the beryllium tile into compression when the assembly is heated up, while the copper / carbon pair puts the carbon edge into tension for the same loading condition. Having the beryllium in compression is favourable compared to tension, since it mean that the first mode of crack opening is suppressed. This is one more justification for a better bonding behaviour of the Be/Cu pair compared to C/Cu, although this assertion could be attenuated by the fact that cyclic loading will distribute tension and compression loading over many heat cycles.

Stress analysis is pursued to investigate possible governing factors to the bond damage. These factors involve bond temperature at design heat load, or the thermal gradient (itself a consequence of the surface heat flux density). This investigation is suggested by the analogy between damage threshold of both EHF and NHF concepts, which both appear to present increasing probability of damage when the Cu/Be bond is operated in a temperature range from 300 to 350°C. An hypothesis appears to make sense : the bond damage trigger is set primarily by its temperature, more than by the shear induced by the thermal gradient. This hypothesis is tested by analysis, using temperature-dependent material properties and kinematic hardening material curves.

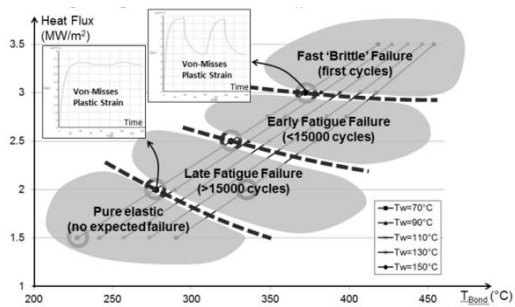

Fig. 4. : Calculated iso-strain curves for NHF fingers

The numerical testing is applied to the NHF design. In the absence of well-established bond damage criteria, the strain range of the base material is adopted as damage indicator. The beryllium to copper bond itself is assumed to be perfectly joined. The strain is picked up in the soft copper which is the weakest material around the bond, far enough from the singularity point so that the calculated strain makes sense. Figure 4 shows iso-strain curves. The figure indicates that the iso-strain depends mainly on the heat flux. The effect of temperature on the iso-strain is modest, especially at the higher heat load of 3 MW/m². Only at the lower heat load of 2 MW/m² does the bond temperature affect the damage level. Overall, the analysis tends to show that the stress caused by the thermal gradient is the dominant cause for damage to the bonding zone. This analysis is purely numerical, and would need to be substantiated by further testing.

## 4. Conclusion

The preparation of the manufacture of the ITER FW panels is now well in the qualification phase. Three procuring agencies are preparing semi-prototypes for qualification, EU, RF and CN. One qualification prototype is being tested under heat flux cycling, while manufacturing of additional prototypes is in progress. In parallel, bond defects program are being done, toward defining acceptable bond defect size during the series manufacturing. The bonding of beryllium tile to copper is observed to be rather resilient to bond defect propagation. Thermal and stress analyses by finite element modelling were done in support of the development and testing activities, toward helping interpretation and validation of test results. The analyses indicate that the stress singularity is considerably less for the beryllium to copper bond, compared to the former fusion experience with carbon to copper bond. However more investigations are required before a Be/Cu bond damage model can be drafted. Hence component qualification on a heat flux test bed remains an essential tool toward bringing component quality to the required level.

## Disclaimer